\documentclass[hyper]{JHEP} % 10pt is ignored!

\usepackage{epsfig}

\newcommand\fverb{\setbox\pippobox=\hbox\bgroup\verb}
\newcommand\fverbdo{\egroup\medskip\noindent%
			\fbox{\unhbox\pippobox}\ }
\newcommand\fverbit{\egroup\item[\fbox{\unhbox\pippobox}]}
\newbox\pippobox
\title{Non-BPS Dp-brane in Dk-Brane Background}

\author{by J. Kluso\v{n}\\
	 Department of Theoretical Physics and Astrophysics\\
                   Faculty of Science, Masaryk University\\
Kotl\'{a}\v{r}sk\'{a} 2, 611 37, Brno\\
Czech Republic\\
	E-mail: \email{klu@physics.muni.cz}}

\preprint{\hepth{}}

\abstract{In this paper we will study the
 dynamics of a non-BPS Dp-brane in 
the background of $N$ BPS Dk-branes.}


\def\tr{\mathrm{Tr}}

\def\bA{\mathbf{A}}

\def\bI{\mathbf{I}}
\begin{document}
%%%%%%%%%%%%%%%%%%%%%
%%%%Introduction %%%%%%%%%
%%%%%%%%%%%%%%%%%%%%
\section{Introduction}\label{first}
One of the most important challenges
in string theory is to formulate the 
framework for addressing the 
string dynamics in time dependent
backgrounds. Due to the 
fundamental work by A. Sen
\cite{Sen:2002qa,Sen:2002in,Sen:2002nu}
there has been significant progress in
the understanding time-dependent 
phenomena in open string theory
\footnote{For recent review, see
\cite{Sen:2004nf} where extensive
list of relevant papers can be found.}: 
decay
of unstable D-brane or brane-antibrane
pair can be described in 
terms of condensation of open string
tachyon. 

Another problem where time dependent
string dynamics is involved is 
situation
when a probe Dp-brane moves 
in the background
of $N$ NS5-branes. 
Very interesting
analysis of 
 this process was given 
by Kutasov in \cite{Kutasov:2004dj} 
\footnote{The extension of this 
approach can be found in
\cite{Nakayama:2004ge,Chen:2004vw,
Thomas:2004cd,Bak:2004tp,Kluson:2004yk,
Kluson:2004xc,Saremi:2004yd,Kutasov:2004ct,
Sahakyan:2004cq,Ghodsi:2004wn,Panigrahi:2004qr,
Yavartanoo:2004wb,Nakayama:2004yx}.}.
It was shown there that the dynamics of
the radial mode of the BPS D-brane
resembles the tachyon rolling dynamics
of unstable D-brane. In particular, 
for appropriate background of
NS5-branes the radion effective action
takes exactly the same functional
form as the tachyon effective action
for unstable D-brane and proposed to
view the radion rolling dynamics as
a sort of ``geometrical'' realisation
of tachyon rolling dynamics on unstable
D-brane. 

An important message from the study
of the real time open string dynamics is
that the tachyon effective action 
Dirac-Born-Infeld (DBI)-like type 
\cite{Sen:1999md,Kluson:2000iy,
Bergshoeff:2000dq,Garousi:2000tr}
captures many aspects of rolling tachyon
solution of string theory.
Some suggestions why the tachyon DBI action is
such 
remarkable successful in the description
of  the tachyon
dynamics were presented in 
\cite{Kutasov:2003er,Niarchos:2004rw,Sen:2003zf}
\footnote{For alternative point of view on the
problem of tachyon effective action, see 
\cite{Fotopoulos:2003yt}.}. 
Then it is certainly useful 
 to gain as  many informations
from the study of the effective field
theory description of
 various D-brane
configurations as possible. For that reason 
we have recently
analysed the effective field theory
description of the non-BPS Dp-brane in
the background of $N$ NS5-branes 
\cite{Kluson:2004yk,
Kluson:2004xc}.  
We have shown that
the tachyon effective action can
be written in the form that  suggests
geometric origin  of the tachyon as the new
embedding coordinate. Then the
tachyon effective action can be considered
as an action for a BPS D-brane 
moving
in nontrivial eleven dimensional background.
This observation supports 
the ideas that were previously
presented 
 in \cite{Kutasov:2004dj,Kutasov:2004ct}.
Then we have studied the dynamics of
the non-BPS Dp-brane in the NS5-brane
background. We have considered two
cases: The first one when the tachyon
and radial mode were time dependent.
We have argued that the general analysis of this
problem     is 
very difficult  
 thanks to the non-existence
of the  additional conservation charge related
to the presence of the tachyon. 
On the other hand in the region of the field
theory space where $T$ is large and
unstable D-brane was close to the NS5-branes
the tachyon effective action posses 
additional symmetry  leading to the 
emergence of the new
 conserved charge. 
Using this fact we were able to determine
the time dependence of
the tachyon and radial mode in this particular
region of the field theory space. 
The second example
of the tachyon condensation
 that we have studied
was the situation when the
tachyon was  function of
one single spatial coordinate on 
the worldvolume of unstable D-brane
while the embedding modes were time dependent.

In this paper we extend the
analysis  performed in \cite{Kluson:2004yk} 
to the 
case of the motion of a non-BPS 
Dp-brane in the
background of $N$ BPS Dk-branes.
The situation when a BPS Dp-brane
moves in the Dk-brane background was studied
previously in \cite{Saremi:2004yd,Panigrahi:2004qr}
and in \cite{Burgess:2003mm}. Even if
the analysis of this problem is slightly
more complicated than
 the study of the dynamics of
a BPS Dp-brane in the NS5-brane background
one can still find an analogue between
the rolling radion and rolling tachyon.
In particular, it was shown 
in \cite{Panigrahi:2004qr} that the 
effective field theory description of the
dynamics of the  BPS Dp-brane in the
background of $N$ Dk-branes can be still
mapped to the  tachyon like effective
action for unstable Dp-brane.  
Then it is natural to
 extend this
 analysis to the case when
 a non-BPS Dp-brane moves in the background
of $N$ Dk-branes.  

More precisely, in the next section (\ref{second}) 
we will write the general form of the non-BPS Dp-brane
effective action in the background of $N$ BPS Dk-branes.
We will show  that the tachyon mode
has striking similarity with additional embedding
coordinate. Then we will focus on the
study of the dynamics of the non-BPS Dp-brane
in the Dk-brane background. 
We will firstly consider
the case of the time dependent tachyon 
 and  time dependent
radial mode. After the general discussion
of the properties of the non-BPS Dp-brane
effective action for this  dynamical
situation  we will
consider in section (\ref{third})
 the case when the
tachyon is large and unstable Dp-brane is near 
 the worldvolume of  the
coincident $N$   Dk-branes. Following 
\cite{Kluson:2004yk} 
 we will try to find
   an additional symmetry of the
tachyon effective action. 
Since the form
of the metric and dilaton produced by $N$ Dk-branes
is slightly more complicated
than 
in the case of the
NS5-brane background we will  need 
to consider more general form of the transformation
of the tachyon and the radial mode.
However the condition that probe
Dp-brane action has to be invariant
under these transformations further
restrict the dimension of
the background Dk-branes, namely we
will see that the spatial dimension
of Dk-branes should be equal to three. 

Then for the  case of the background
of $N$ D3-branes 
we will be able to determine the time
dependence of the tachyon and radial
mode. We will also find
that the time dependence of the
radial mode and the tachyon depends
on the spatial dimension of the
 worldvolume of unstable Dp-brane.
 
As the second example of the tachyon
dynamics we will consider in section
(\ref{fourth}) the case
when the tachyon is spatial dependent
and the radial mode is time dependent. 
 We will see that the tachyon
condensation results to the emergence
of a singular tachyon kink that has 
natural interpretation as a D(p-1)-brane that
moves in the background of $N$ 
Dk-branes. Then we will argue
that the description of 
 the dynamics of the resulting
configuration can be performed 
in the same way as in  
\cite{Burgess:2003mm,Panigrahi:2004qr}.

Finally, in conclusion (\ref{fifth})
we outline our result and suggest 
possible extension of this work. 
%%%%%%%%%%%%%%%%%%%%%%%%%%%%%%
%%%calculation            %%%%
%%%%%%%%%%%%%%%%%%%%%%%%%%%%%%
\section{The effective action for non-BPS
Dp-brane in  Dk-brane background}\label{second}
In this section we will analyse the motion of a
non-BPS Dp-brane in the stack of 
coincident and static Dk-branes 
using the tachyon effective action proposed in
\cite{Sen:1999md,
Bergshoeff:2000dq,Garousi:2000tr,Kluson:2000iy}.
The metric, the dilaton $(\Phi)$, and the R-R
field (C) for  a system of $N$ coincident 
Dk-branes is given by
\begin{eqnarray}\label{Dkbac}
g_{\alpha \beta}=H_k^{-\frac{1}{2}}\eta_{\alpha\beta}
\ , g_{mn}=H_k^{\frac{1}{2}}\delta_{mn} \ ,
(\alpha, \beta=0,1,\dots,k \ , 
m, n=k+1,\dots,9) \ , \nonumber \\
e^{2\Phi}=H_k^{\frac{3-k}{2}} \ , 
C_{0\dots k}=H_k^{-1} \ , H_k=1+\frac{N g_s}{r^{7-k}} 
\ , \nonumber \\
\end{eqnarray}
where $H_k$ is a harmonic function of $N$ 
Dk-branes satisfying the Green function equation in
the transverse space. 

Now let us consider
a non BPS Dp-brane with $p<k$ 
that is inserted
in the background (\ref{Dkbac}) with its
spatial section   stretched
in directions $(x^1,\dots,x^p)$. We will  label
the worldvolume of the non-BPS 
Dp-brane by $\xi^{\mu} ,
\mu=0,\dots,p$ and use reparametrisation invariance
of the worldvolume of the Dp-brane 
to set $\xi^{\mu}=x^{\mu}$.
The position of 
the D-brane in the transverse directions $(x^{p+1},\dots,x^9)$
gives to rise  to scalar fields on the worldvolume
of D-brane, $(X^{p+1}(\xi^{\mu}),\dots X^9(\xi^{\mu}))$. 
According to \cite{Sen:1999md,
Bergshoeff:2000dq,Garousi:2000tr,Kluson:2000iy}
     the non-BPS 
Dp-brane (DBI)-like effective action
takes the form
\begin{equation}\label{actnon}
S=-\tau_p\int  d^{p+1}\xi V(T)e^{-\Phi-\Phi_0}
\sqrt{-\det(G_{\mu\nu}+B_{\mu\nu}+
\partial_{\mu}T\partial_{\nu}T)}+S_{WZ} \ ,
\end{equation}
where $\tau_p=\frac{\sqrt{2}}{
(2\pi)^{p+1}}$ is a tension of the non-BPS Dp-brane
\footnote{We work in units where $
\alpha'=1$.} and 
where $V(T)$ is a tachyon potential. 
According to papers 
\cite{Lambert:2003zr,Kutasov:2003er}
we presume that it takes the form
\begin{equation}
V(T)=\frac{1}{\cosh\frac{T}{\sqrt{2}}} \ . 
\end{equation}
 Finally, in (\ref{actnon}) we have
also included the coupling of 
the non-BPS Dp-brane  to
the Ramond-Ramond field that
is governed by 
 the Wess-Zumino term  
\cite{Okuyama:2003wm,Billo:1999tv}
\begin{equation}\label{Swz}
S_{WZ}=\frac{1}{\sqrt{2\pi}}
\int V(T)dT\wedge C_{RR} \ . 
\end{equation}
Even if the action (\ref{actnon}) presented
above is the most familiar
one that is used for the description
of a unstable D-brane we will prefer 
to describe the  
dynamics of
the non-BPS Dp-brane  
using the tachyon effective
action  
 proposed in \cite{Kutasov:2003er} 
that in the flat spacetime 
takes the form
\begin{eqnarray}\label{Kutact}
S=-\tau_p\int d^{p+1}\xi e^{-\Phi_0} \frac{1}{
\sqrt{1+\frac{T^2}{2}}}
\sqrt{-\det\left(
\eta_{\mu\nu}+
(1+\frac{T^2}{2})^{-1}\partial_{\mu}T\partial_{\nu}T
\right)}+S_{WZ}=\nonumber \\
=-\tau_p\int d^{p+1}\xi
\sqrt{F}e^{-\Phi_0}\sqrt{-\det(\eta_{\mu\nu}
+F\partial_{\mu}T\partial_{\nu}T)}+S_{WZ}
 \ ,
F=\frac{1}{1+\frac{T^2}{2}} \ . 
\nonumber \\
\end{eqnarray}
It can be shown \cite{Kutasov:2003er} that  
there exists field redefinition that maps
(\ref{Kutact}) to (\ref{actnon}).
The existence of this field redefinition implies
that these two  actions are equivalent, at least
at the flat spacetime.

As a next step we will presume  that
  the action (\ref{Kutact})
correctly describes the dynamics of
the non-BPS Dp-brane in the general
closed string background where the
action (\ref{Kutact}) takes the form
\begin{equation}\label{actng}
S=-\tau_p\int d^{p+1}\xi
e^{-\Phi}\sqrt{F}\sqrt{-\det(G_{\mu\nu}
+F\partial_{\mu}T\partial_{\nu}T)}+S_{WZ}
 \ ,
\end{equation}
where the determinant in
(\ref{actnon}) runs over the worldvolume 
directions $\mu=0,\dots,p$, $G_{\mu\nu}$ and 
$B_{\mu\nu}$ are induced metric and $B$ field
on the non-BPS Dp-brane
\begin{eqnarray}
G_{\mu\nu}=\frac{\partial X^A}{\partial \xi^{\mu}}
\frac{\partial X^B}{\partial \xi^{\nu}}G_{AB}(X) \ ,
\nonumber \\
B_{\mu\nu}=\frac{\partial X^A}{\partial \xi^{\mu}}
\frac{\partial X^B}{\partial \xi^{\nu}}B_{AB}(X) \ . 
\nonumber \\
\end{eqnarray}
In this notation
 the indices $A,B=0,\dots,9$ 
run over the whole ten 
dimensional spacetime so that $G_{AB}$ and
$B_{AB}$ are metric and $B$-field in ten dimensions.
It is however important to stress that the
step from (\ref{Kutact}) to (\ref{actng}) 
is  nontrivial since it is well known
that the form of the tachyon effective action
strongly depends on the field theory space where
it is presumed to be valid 
\cite{Fotopoulos:2003yt}.

Let us now  consider 
the non-BPS Dp-brane in the 
background of $N$ Dk-branes 
given in (\ref{Dkbac})
 where the  worldvolume of
the unstable Dp-brane is stretched
along the worldvolume of  Dk-branes.
If we choose the  static gauge 
$\xi^{\mu}=x^{\mu}, \ \mu=0,1,\dots,p$
then  the action (\ref{actng}) 
takes the form 
\begin{eqnarray}\label{actng2}
S=-\tau_p\int d^{p+1}\xi
\sqrt{F}H_k^{\frac{k-3}{4}}
\sqrt{-\det(G_{\mu\nu}
+F\partial_{\mu}T\partial_{\nu}T)}
+S_{WZ} \ , \nonumber \\ 
\end{eqnarray}
where  
the induced metric
is equal to
\begin{equation}
G_{\mu\nu}=H_k^{-\frac{1}{2}}\eta_{\mu\nu}+
H_k^{-\frac{1}{2}}
\partial_{\mu}Y^i\partial
_{\nu}Y^i+
\partial_{\mu}X^m\partial_{\nu}X^m
 H_k^{\frac{1}{2}}
\ . 
\end{equation}
From here on we reserve 
the coordinates
$X^m$ with ($m=k+1,\dots,9$) 
for dimensions
transverse to Dk-brane and the 
coordinates $Y^i$ with ($i=p+1,\dots,k$) 
for those dimensions 
transverse
to non-BPS Dp-brane, but parallel
to the Dk-brane. 

Looking at the form of the tachyon
effective action (\ref{actng2}) 
we can now  think about 
it 
 as the DBI action for
Dp-brane embedded in eleven dimensional
spacetime with the background metric
and dilaton 
\begin{eqnarray}
ds^2=H^{-\frac{1}{2}}
\eta_{\alpha\beta}dx^{\alpha}dx^{\beta}
+H^{\frac{1}{2}}\delta_{mn}dx^mdx^n+FdT^2 \ , 
e^{2\Phi}=\frac{H_k^{\frac{3-k}{2}}}{F}
 \ .
\nonumber \\
\end{eqnarray}
Even if this is clearly very promising
idea it remains to be seen whether it
has more precise physical justification.

Now we will study  the equation of motion
that arises from (\ref{actng2}).
As the first example we will consider the
case when the tachyon and 
the embedding fields $Y^i,X^m$ are time dependent.
 In this
case the matrix 
\begin{equation}
\bA_{\mu\nu}\equiv H_k^{-\frac{1}{2}}\eta_{\mu\nu}
+H_k^{-\frac{1}{2}}\partial_{\mu}Y^i
\partial_{\nu}Y^i
+H_k^{\frac{1}{2}}\partial_{\mu}X^m\partial_{\nu}X^m
+F\partial_{\mu}T\partial_{\nu}T
\end{equation}
takes the form
\begin{equation}
\bA_{\mu\nu}=
\left(\begin{array}{cc}
-H_k^{-\frac{1}{2}}+
H_k^{-\frac{1}{2}}\dot{Y}^i
\dot{Y}^i+
H^{\frac{1}{2}}_k\dot{X}^m\dot{X}^m
+F\dot{T}^2 & 0 \\
0 & H^{-\frac{1}{2}}_k\bI_{p\times p} \\
\end{array}\right)
\end{equation}
so that
\begin{equation}
\det \bA=-H_k^{-\frac{1}{2}}(1-
\dot{Y}^i\dot{Y}^i-
H_k\dot{X}^m\dot{X}^m
-H_k^{\frac{1}{2}}F\dot{T}^2)H_k^{-\frac{p}{2}} \ . 
\end{equation}
Then the action (\ref{actng2}) takes the form
\begin{equation}\label{aTR}
S=-\tau_p V_p\int
dt \sqrt{F}H_k^{\frac{k-p-4}{4}}
\sqrt{1-\dot{Y}^i
\dot{Y}^i-H_k\dot{X}^m\dot{X}^m-H_k^{\frac{1}{2}}
F\dot{T}^2}\equiv -\int dt \mathcal{L} \ . 
\end{equation}
Note that for 
$T=0$ the action
(\ref{aTR}) agrees with 
the action studied
in \cite{Burgess:2003mm}. 
We can also see 
 that for the time dependent tachyon the
WZ term vanishes identically using the
fact that $dT=\dot{T}dt$ and that
the nonzero components of the RR fields 
are these with the factor $C_{01\dots p}$ so
that the wedge product $dT\wedge C$ vanishes.  

To proceed  we now determine 
some conserved
charges that follow from  (\ref{aTR}).
Since the action does not explicitly depend on 
time  the Hamiltonian 
\begin{eqnarray}\label{Ham}
H=P_i\dot{Y}^i+P_m\dot{X}^m+\Pi\dot{T}-\mathcal{L}=
\nonumber \\
=\frac{\tau_pV_p\sqrt{F}
H_k^{\frac{k-p-4}{4}}}{
\sqrt{1-\dot{Y}^i
\dot{Y}^i-H_k\dot{X}^m\dot{X}^m-
H_k^{\frac{1}{2}}
F\dot{T}^2}} \ 
\end{eqnarray}
is conserved. In 
(\ref{Ham}) 
the momenta conjugate
to $T,X^m$ and $Y^i$ 
are equal to  
\begin{eqnarray}
P_m=\frac{\delta \mathcal{L}}{\delta \dot{X}^m} =
\tau_pV_p H_k^{\frac{k-p-4}{4}}
\frac{H_k \dot{X}^m}{\sqrt{1-\dot{Y}^i\dot{Y}^i-
H_k\dot{X}^m\dot{X}^m-
H_k^{\frac{1}{2}}
F\dot{T}^2}} \ , 
\nonumber \\
P_T=\frac{\delta \mathcal{L}}
{\delta \dot{T}}=\tau_pV_pH_k^{\frac{k-p-4}{4}}
\frac{H^{\frac{1}{2}}F\dot{T}}{
\sqrt{1-\dot{Y}^i
\dot{Y}^i-H_k\dot{X}^m\dot{X}^m-
H_k^{\frac{1}{2}}
F\dot{T}^2}} \ ,
\nonumber \\
P_i=\frac{\delta \mathcal{L}}
{\delta \dot{Y}^i}=\tau_pV_p H_k^{\frac{k-p-4}{4}}
\frac{\dot{Y}^i}{\sqrt{1-\dot{Y}^i\dot{Y}^i-
H_k\dot{X}^m\dot{X}^m-
H_k^{\frac{1}{2}}
F\dot{T}^2}} \ . 
\nonumber \\
\end{eqnarray}
The next
 conserved charge corresponds to the 
manifest rotation
symmetry $SO(9-k)$ 
of the transverse $R^{9-k}$ space. 
In fact, it easy to see 
that the action is invariant under
the transformation 
\begin{equation}
X'^m(t)=\Lambda^m_nX^n(t) \ , 
\Lambda^m_k\delta_{mn}
\Lambda^n_l=\delta_{kl} \ . 
\end{equation}
Using this  symmetry we can restrict
ourselves to the motion in the transverse
$(x^8,x^9)$ plane
\footnote{This restriction also implies
that the background Dk-branes are those
with $k<8$.} where the corresponding
conserved angular momentum is equal to
\begin{eqnarray}
L\equiv L^9_8=
\tau_pV_p\sqrt{F}H_k^{\frac{k-p}{4}}
\frac{\dot{X}^9X^8-
\dot{X}^8X^9}
{\sqrt{1-\dot{Y}^i\dot{Y}^i-
H_k\dot{X}^m\dot{X}^m-H_k^{\frac{1}{2}}
F\dot{T}^2}}=
P^8X^9-P^9X^8 \ . 
 \nonumber \\
\end{eqnarray}
In terms of the 
polar coordinates defined as
\begin{equation}
X^8=R\cos \theta \ ,
X^9=R\sin \theta \ 
\end{equation}
the conserved energy 
and angular momentum
take the form
\begin{eqnarray}\label{conq}
E=\tau_pV_p\sqrt{F}
H_k^{\frac{k-p-4}{4}}
\frac{1}{
\sqrt{1-\dot{Y}^i\dot{Y}^i-
H_k(
\dot{R}^2+R^2\dot{\theta}^2)
-H_k^{\frac{1}{2}}
F\dot{T}^2}} \ , \nonumber \\
L=\tau_pV_p\sqrt{F}
H_k^{
\frac{k-p}{4}}
\frac{R^2\dot{\theta}}
{\sqrt{1-\dot{Y}^i
\dot{Y}^i-H_k(\dot{R}^2+
R^2\dot{\theta}^2)-H_k^{\frac{1}{2}}
F\dot{T}^2}} \ .  \nonumber \\
\end{eqnarray}
It is also convenient to work with the
densities so that we strip 
off the volume
factor $V_p$ in 
(\ref{conq}). Then in what
follows  we will consider
 $E,L$ as
corresponding densities. 

Finally, since the action 
(\ref{aTR}) does not explicitly
depend on $Y^i$ it is clear that
the momenta $P_i$ are conserved
as well. In order to simplify
expressions that we will obtain
below   we will consider
the case when all   momenta
$P_i$ vanish and hence $\dot{Y}^i=0$.

As can be seen from the form
of the tachyon effective action (\ref{aTR}) 
and corresponding conserved charges
it is very difficult to solve the resulting
differential equations for $R,T$ and $\theta$ in the
full generality. Moreover, the fact that there is 
missing the third conserved charge makes the
analysis even worse. On the other hand we have
shown in \cite{Kluson:2004yk} that for the
case of the motion of a non-BPS Dp-brane
in the background of $N$ NS5-branes one
can find the region in the field theory
space spanned by $T,R$, where new additional symmetry
of the non-BPS Dp-brane action emerges. 
The existence of this symmetry allows to define
new conserved charge and one can explicitly
describe the dynamics of the non-BPS Dp-brane.
For that reason it would be desirable to find
such a symmetry for non-BPS Dp-brane in
the Dk-brane background as well. 

Before we address this question  we
would like to say  few words about
the dynamics of non-BPS Dp-brane in
the Dk-background, 
following \cite{Burgess:2003qv,
Burgess:2003mm}.
For convenience we again
 write the action that governs
the dynamics of a non-BPS Dp-brane in
the $(x^8,x^9)$ plane 
\begin{equation}\label{aTG}
S=-\tau_pV_p\int dt H_k^{\frac{k-p-4}{4}}
\sqrt{F}\sqrt{1-H_k(\dot{R}^2+R^2\dot{\theta}^2)
-FH_k^{\frac{1}{2}}F\dot{T}^2} \ .
\end{equation}
From this action we immediately
 obtain the 
momenta $P_R,P_T$ and $P_{\theta}$
\begin{eqnarray}
P_R=
\tau_pV_p dt H_k^{\frac{k-p-4}{4}}
\sqrt{F}\frac{H_k\dot{R}}
{\sqrt{1-H_k(\dot{R}^2+R^2\dot{\theta}^2)
-FH_k^{\frac{1}{2}}F\dot{T}^2}}
 \ ,
\nonumber \\
P_{\theta}=
\tau_pV_p H_k^{\frac{k-p-4}{4}}
\sqrt{F}\frac{R^2H_k\dot{\theta}}
{\sqrt{1-H_k(\dot{R}^2+R^2\dot{\theta}^2)
-FH_k^{\frac{1}{2}}F\dot{T}^2}} \ , 
\nonumber \\
P_T=
\tau_pV_p H_k^{\frac{k-p-4}{4}}
\sqrt{F}\frac{FH_k^{\frac{1}{2}}\dot{T}}
{\sqrt{1-H_k(\dot{R}^2+R^2\dot{\theta}^2)
-FH_k^{\frac{1}{2}}F\dot{T}^2}} \ .
\nonumber \\
\end{eqnarray}
Hence the Hamiltonian density 
is equal to
\begin{eqnarray}
H=\frac{1}{V_p}
\left(P_R\dot{R}+P_{\theta}\dot{\theta}+
P_T\dot{T}-\mathcal{L}\right)
=\sqrt{\tau_p^2 H_k^{\frac{k-p-4}{2}}F+
\frac{P^2_R}{
H_k}+\frac{P^2_{\theta}}{R^2H_k}+\frac{P^2_T}
{FH_k^{\frac{1}{2}}}} \ .  \nonumber \\
\end{eqnarray}
Since the Hamiltonian is 
monotonically
increasing function of canonical momenta
it is bounded from bellow 
by the potential
\begin{equation}
V(T,R)\equiv H(P_R=P_T=P_\theta=0)
=\tau_p H_k^{\frac{k-p-4}{4}}\sqrt{F} \ .
 \end{equation}
 For fixed $T$ this potential is
attractive for $k-p<4$ and repulsive
when $k-p>4$. 

We also see that the Lagrangian
in (\ref{aTG})
does not explicitly depend on $\theta$
and consequently 
 the momentum
$P_{\theta}$ is conserved. This fact
allows us to introduce 
the effective potential for $T,R$ 
that has the form
\begin{equation}\label{Ve}
V_{eff}(T,R)\equiv H(P_R=P_T=0)
=
\sqrt{\tau^2_p H_k^{\frac{k-p-4}{2}}F+
\frac{P_{\theta}^2}{H_kR^2}} \ . 
\end{equation}
 The utility
of this potential follows 
from the fact that 
$\frac{\partial E}{\partial P_{R,T}}>0$. 
Then the
allowed ranges of $ R,T$
 can be found by
standard device by plotting 
$V_{eff}$ against 
$T,R$ and finding those $T, R$ where
$E>V_{eff}(R,T)$. 

In fact, our goal is to find  the stable orbits 
corresponding
to the extremum of the potential. 
The possible extrema of   $V_{eff}$
 with respect
to $T$ follow from the
equation 
\begin{equation}\label{ved}
0=\frac{\delta V_{eff}(T,R)}
{\delta T}=-
\frac{\tau_pH_k^{\frac{k-p-4}{2}}}
{2\sqrt{\tau^2_p H_k^{\frac{k-p-4}{2}}F+
\frac{P_{\theta}^2}{H_kR^2}}}
\frac{T}{
(1+\frac{T^2}{2})^2} \ . 
\end{equation}
The first extremum corresponds 
to 
 $T=0$ with the standard
interpretation as the unperturbed
 unstable Dp-brane. The
 analysis of 
 the dynamics of non-Dp-brane 
sitting at its unstable point $T=0$ 
is similar to the analysis performed in 
\cite{Burgess:2003qv,Burgess:2003mm}
as we will review bellow. 

As can be seen from (\ref{ved}) the stable extrema
of the effective potential occur at $T=\pm\infty$. 
It is believed that these extrema 
correspond to some form of the pressureless
gas. 
It would be certainly interesting
to study the dynamics of this mysterious
form of matter at the vicinity
of the Dk-branes and we hope to return
to this problem in future. 

Let us now consider the case when the
tachyon sits in its unstable
minimum $T=0$ and study the 
properties of the effective potential
(\ref{Ve}) for small and large $R$. 
For $R\rightarrow 0$ the effective
potential (\ref{Ve}) is equal to
\begin{equation}\label{Vers}
V_{eff}(R,T=0)=
\sqrt{\tau_p^2(Ng_s)^{\frac{(k-p-4)}{2}}
R^{\frac{(k-7)(k-p-4)}{2}}+
\frac{P^2_{\theta}R^
{5-k}}{Ng_s}} \ .
\end{equation}
Since we presume that $k<7$ the sign
of the first exponent in (\ref{Vers})
is given by sign of $k-p-4$. 
Then we can find for small $R$ following
limits:
\begin{itemize}
\item For $k=6$ the last term in
(\ref{Vers}) diverges ensuring
 that the potential 
diverges at origin.
\item For $k=5$ the last term 
approaches
the constant for small $r$. 
Then for $p=4,2$ the potential 
approaches the
constant $\frac{P_{\theta}}{\sqrt{Ng_s}}$ 
in this limit. On the other hand for
$p=0$ the potential diverges as
$V_{eff}\sim \frac{1}{\sqrt{R}}$. 
\item For $k=4$ the last term vanishes
at the origin and the potential 
converges to zero since the first
term scales as $R^{3p}$ for $p=3,1$. 
\item For $k=3$ the last term again
vanishes at the origin. The first
term scales as $R^{2(p+1)}$ that converges
to  zero for $p=0,2$.
\item  For $k=2$ the last term again
vanishes at the origin and the potential
vanishes since the first term scales
as $R^{\frac{5(2+p)}{2}}$. 
\item Finally, for $k=1$ the last term
vanishes and the first term scales as
$R^{p(p+3)}$.
\end{itemize}
On the other hand, 
in the large $R$
limit we instead find
\begin{equation}
V_{eff}(R,T=0)=\sqrt{\tau_p^2
\left(1+\frac{(k-p-4)Ng_s}{2R^{7-k}}
\right)
+\frac{P^2_{\theta}}{R^2}} \ 
\end{equation}
and we see that this potential approaches
the constant for 
$R\rightarrow \infty$. 
How this limit is approaching 
depends on $p,k$ in the following ways:
\begin{itemize}
\item For $k=6$ the second term dominates
and the potential approaches its limit
from bellow.
\item For $k=5$ the way the potential
approaches its limiting value
depends on $p$. 
For $p=0$ we have
$\frac{\tau_p^2Ng_s(k-p-4)}{2}+
P_{\theta}^2=\frac{\tau_p^2Ng_s}{2}+
P^2_{\theta}>0$ and the
potential is reached from above. 
On the other hand for $p=2,4$
the way the potential approaches the
limiting value depends on the
sign of 
$\frac{\tau_p^2Ng_s(k-p-4)}{2}+P_{\theta}^2
=\frac{\tau_p^2Ng_s(1-p)}{2}+P_{\theta}^2$.
\item For $k<5$ the last term always
dominates and the potential approaches its
limit from above.  
\end{itemize}
Combining the behaviour for
small and large $R$ we obtain
following picture:
\begin{itemize}
\item For $k=6$ the $V_{eff}$ reaches
the minimum away from $R=0$ and 
bound orbit exist. The analysis of
this situation is the same as in the
case of BPS Dp-brane studied in
\cite{Burgess:2003mm}. It is also
clear that these bound orbits are 
unstable since the tachyon is
sitting in its unstable maximum $T=0$.
Note also that as opposite to the
case of BPS Dp-brane 
there cannot exist
the supersymmetric state characterised 
by condition $k-p=4$ 
since for  unstable Dp-brane probe
$k-p$ is always odd number.
\item If $k=5$ and $p=2,4$ the existence
of minimum depends on $P_{\theta}$ and there
is a bound orbit if $\frac{\tau_p^2Ng_s
(p-1)}{2}<P^2_{\theta}$. 
For $p=0$ there are not bound orbits.
\item For $k<5$ the potential has the
minimum at the origin that is smaller
than at infinity. This implies 
that localised
orbits exist.  
\end{itemize}
Once again we must stress that
the picture outlined above corresponds
to the situations when
$T=0$. On the other hand
 we will rather consider
the case when the tachyon
is dynamical as well. 
In fact, since the analysis
of an unstable D-brane with the
vanishing tachyon is almost the
same as the analysis of the
dynamics of BPS Dp-brane in Dk-brane
background
that was studied very carefully in \cite{Burgess:2003mm}
we will not discuss it here further. 
%%%%%%%%%%%%%%%%%%%%%%%%%%%%%%%%
%%%large T and small R                      %%%%%%%
%%%%%%%%%%%%%%%%%%%%%%%%%%%%%%%
\section{Large T and small R}\label{third}
We have shown in
 our previous paper 
\cite{Kluson:2004xc}
where the dynamics of non-BPS
Dp-brane in the background of
NS5-branes was studied that 
 for large value of tachyon  and
in the region close to the
worldvolume of $N$ NS5-branes
a non-BPS D-brane
action has additional symmetry
that considerably simplifies the
analysis of the time evolution
of the non-BPS Dp-brane. Then
it is natural to ask the question
whether similar symmetry emerges
in  case when  the
 non-BPS Dp-brane is embedded in
the background of $N$ Dk-branes
in the region of the field theory
space when  
$\frac{\lambda}{R^{7-k}}\gg 1, \ 
\lambda\equiv Ng_s \ ,\frac{T^2}{2}\gg 1$. 
In this region of the field theory
space spanned by $T,R$ the action 
(\ref{aTR}) takes the form
\begin{equation}\label{aTRl}
S=-\tau_p \sqrt{2}V_p
\lambda^{\frac{k-p-4}{4}}
\int dt \frac{1}{TR^{\frac{(7-k)(k-p-4)}{4}}}
\sqrt{1-\frac{\lambda}{R^{7-k}}(\dot{R}^2
+R^2\dot{\theta}^2)-\frac{
2\sqrt{\lambda}}{T^2R^{\frac{7-k}{2}}}
\dot{T}^2} \ . 
\end{equation}
Let us now demand  that 
the action
(\ref{aTRl}) should be
 invariant 
under  following transformations
\begin{equation}
t'=\Lambda^{\alpha}t \ ,
T'(t')=\Lambda^{\beta}T(t)\ , 
R'(t')=\Lambda^{\gamma}R(t) \ ,
\theta'(t')=\Lambda^{\delta}\theta(t) \ . \ 
\end{equation}
Now the requirement that the 
transformed
and original action (\ref{aTRl}) should
be equal leads to the  set of  
following conditions 
\begin{eqnarray}\label{dt}
\alpha-\beta-\frac{(7-k)(k-p-4)}{4}\gamma=0
\ , \nonumber \\
(k-7)\gamma+2\gamma-2\alpha=0 \Rightarrow
\alpha=\frac{(k-5)\gamma}{2} \ , 
\nonumber \\
(k-7)\gamma+2\gamma+2\delta-2\alpha=0
\Rightarrow \delta=0 \ , \nonumber \\
\frac{k-7}{2}\gamma-2\alpha=0
\Rightarrow (k-7)\gamma=4\alpha \ . \nonumber \\
\end{eqnarray}
If we combine   these equations 
then we obtain following
result
\begin{equation}
\gamma=-\alpha \Rightarrow
(k-3)\alpha=0 \  
\end{equation}
with the solution that
 $k=3$ and $\alpha$ is 
arbitrary or 
$k\neq 3$ and $\alpha=0$.
 The second solution
implies  $\gamma=\delta=0$ 
and hence there is not any
symmetry at all. 
More interesting is the
first case.
 Then from 
 the second equation
in (\ref{dt}) we obtain
\begin{equation}
\alpha-\beta-\frac{(7-k)(k-p-4)}{4}\gamma=0
\Rightarrow
 \beta=-p\alpha  \ ,
\end{equation}
where now   $\alpha$ is arbitrary.
In fact, different values of $\alpha$ 
correspond to different values of 
 $\Lambda$ since
\begin{equation}
t'=\Lambda^{\alpha}t=
\tilde{\Lambda}t \ .
\end{equation}
Then we can fix its value to
be equal to an arbitrary
number and we  choose 
 $\alpha=-1$.
To recapitulate, 
 the  transformation 
under which the non-BPS Dp-brane action
(\ref{aTRl}) in the background of
$N$ D3-branes
\begin{equation}\label{aTRk3}
S=-\frac{\tau_p\sqrt{2}V_p}{
\lambda^{\frac{p+1}{4}}}\int
dt \frac{R^{p+1}}{T}
\sqrt{1-\frac{\lambda}{R^{4}}(\dot{R}^2
+R^2\dot{\theta}^2)-\frac{
2\sqrt{\lambda}}{T^2R^{2}}\dot{T}^2}\equiv
-\int dt\mathcal{L} \ . 
\end{equation}
 is invariant takes the form
\begin{equation}\label{s3}
t'=\lambda^{-1}t \ , 
R'(t')=\lambda R(t) \ , 
\theta'(t')=\theta(t) \ , 
T'(t')=\lambda^{p}T(t) \ . 
\end{equation}
The conserved  charge
that generates these transformations
takes the form 
\begin{eqnarray}\label{Dd}
D=-tH-pTP_T-
RP_R  \ . \nonumber \\
\end{eqnarray}
Now we are ready to study 
 the time evolution of a non-BPS
Dp-brane in the background of 
$N$ D3-branes
for large $T$ and small $R$. 
It turns out that it is   convenient to
work in the Hamiltonian formalism.
The Hamiltonian 
that follows from (\ref{aTRk3})
takes the form 
\begin{eqnarray}\label{HamRT}
H=P_R\dot{R}+P_{\theta}\dot{\theta}+
P_T\dot{T}-\mathcal{L}
=\sqrt{
\frac{2\tau_p^2R^{2(p+1)}}{\lambda^{\frac{p+1}{2}}
T^2}+\frac{P^2_RR^4}{\lambda}
+\frac{P^2_{\theta}R^2}{\lambda}
+\frac{P^2_TT^2R^2}{2\sqrt{\lambda}}} \ .
\nonumber \\
\end{eqnarray}
Since the  Hamiltonian (\ref{HamRT})
does not explicitly depend on 
$\theta$ we
get  that $P_{\theta}$ is 
conserved
\begin{equation}
\dot{P}_{\theta}=-\frac{\partial H}
{\delta \theta}=0  \ .
\end{equation}
On the other hand the
time dependence 
of $T$ and $R$ 
is determined by the equations
of motion that follow
from  (\ref{HamRT})  
\begin{eqnarray}\label{dRT}
\dot{R}=\frac{\partial H}{\partial
P_R}=\frac{P_RR^4}{\lambda
\sqrt{
\frac{2\tau_p^2R^{2(p+1)}}{\lambda^{\frac{p+1}{2}}
T^2}+\frac{P^2_RR^4}{\lambda}
+\frac{P^2_{\theta}R^2}{\lambda}
+\frac{P^2_TT^2R^2}{2\sqrt{\lambda}}}}=
\frac{P_RR^4}{\lambda E} \nonumber \\
\dot{T}=\frac{\partial H}{\partial P_T}=
 \frac{P_TT^2R^2}
{2\sqrt{\lambda}\sqrt{
\frac{2\tau_p^2R^{2(p+1)}}{\lambda^{\frac{p+1}{2}}
T^2}+\frac{P^2_RR^4}{\lambda}
+\frac{P^2_{\theta}R^2}{\lambda}
+\frac{P^2_TT^2R^2}{2\sqrt{\lambda}}}}=
\frac{P_TT^2R^2}{2\sqrt{\lambda}E} \ .
\nonumber \\
\end{eqnarray}
Now we use the charge
(\ref{Dd}) to express $P_R$ as
function of $E$ and $R$.
Firstly we will presume  that
various terms in (\ref{Dd}) contribute 
in the following way
\begin{eqnarray}\label{DETR}
-tE-pTP_T=0 \ , 
D=-P_RR  \ .\nonumber \\
\end{eqnarray}
Then the  second equation
in (\ref{DETR}) implies
\begin{equation}
P_R=-\frac{D}{R} \ .
\end{equation}
Inserting this relation into
(\ref{dRT})  we get the
differential equation for $R$
\begin{eqnarray}
\dot{R}=-\frac{D}{\lambda E}R^3
\nonumber \\
\end{eqnarray}
that has the general solution
\begin{equation}
\frac{2Dt}{E}=
\frac{\lambda}{R^2}-\frac{\lambda}{R_0^2}
\ ,
\end{equation}
where we have chosen the initial
condition that at $t_0=0$ the 
Dp-brane is at the point $\frac{\lambda}{R_0^2}\gg 1$. 
The requirement that Dp-brane should move
towards to the worldvolume of $N$ D3-branes
implies that $D>0$. 

To find the time dependence of $T$ we
 use (\ref{DETR}) to express
$P_T$ as a  function of $T$ and $E$  and
insert this expression 
into the second 
equation in (\ref{dRT})
with the result
\begin{eqnarray}\label{dotT}
\dot{T}=-\frac{tET^2R^2}{2pT
\sqrt{\lambda}E}=
-\frac{tT}{2p
\sqrt{\lambda}}\frac{\lambda}{
\frac{2Dt}{E}+\frac{\lambda}{R_0^2}}
\Rightarrow
\nonumber \\
\frac{dT}{T}=
-\frac{R_0^2 t}{2p\lambda^{3/2}}
\frac{1}{1+\frac{2DR_0^2}{\lambda E}
t}=\nonumber \\
\ln \frac{T}{T_0}=
-\frac{E}{4p\sqrt{\lambda}}
t+\frac{E^2\sqrt{\lambda}}
{8pD^2R_0^2}\ln
\left(\frac{2R_0^2D}{\lambda E}
t+1\right) \ .
\nonumber \\
\end{eqnarray}
Since $\frac{2R_0^2D}{\lambda E}
t\approx \mathcal{O}(1)$ we see
that dominant contribution in the 
regime $T\gg 1$ comes from the
term linear in $t$. Then
 in the
first approximation we 
obtain the result 
\begin{equation}
T=T_0e^{-\frac{E}
{4p\sqrt{\lambda}D}t} \ ,
T_0\gg 1 \ .
\end{equation}
We see that the tachyon is
lowering its value from the 
initial large $T_0$ until it reach
the region where the approximation
of large tachyon is not valid. 
On the other hand in the region when
we can trust this solution we can give
the physical interpretation of this
solution as follows: It describes
the initial stage of the brane
creation in which the unstable
Dp-brane evolves from the
state very close to the vacuum
since by presumption $T_0\ll 1$.
At the same time this unstable
D-brane  moves
towards to the worldvolume of
D3-branes.  

As the second example that can
be easily solved  we take the
ansatz when $D$ splits as
\begin{equation}\label{DTRs}
D=-pTP_T \ , 
tE=-P_RR \ . 
\end{equation}
Now the differential equation for $R$ is
\begin{equation}
\dot{R}=-\frac{tR^3}{\lambda}
\end{equation}
that has the solution
\begin{equation}\label{Rt1}
\frac{\lambda}{R^2}=t^2+\frac{\lambda}{R_0^2} \ .
\end{equation}
This solution  again describes Dp-brane
moving towards to the worldvolume of D3-branes.
On the other hand the differential equation
for $T$, using (\ref{DTRs}) and
also  (\ref{Rt1}) takes the form
\begin{eqnarray}
\dot{R}=-\frac{D}{2\sqrt{\lambda}pE}TR^2
\Rightarrow \nonumber \\
\frac{dT}{T}=-\frac{D\sqrt{\lambda}}
{2Ep}\frac{1}{t^2+\frac{\lambda}{R_0^2}} 
\nonumber \\
\end{eqnarray}
with the solution
\begin{equation}\label{Tttan}
T=T_0e^{-\frac{DR_0}{2Ep}
\arctan\frac{R_0^2t^2}{\lambda}} \ .
\end{equation}
Since by presumption $\frac{R_0^2}{\lambda}\ll 1$ 
it is reasonable for $t^2\ll \frac{\lambda}{R_0^2}$ 
to approximate $\arctan \frac{R_0^2t^2}{\lambda}$
with $\frac{R_0^2t^2}{\lambda}$. Then
 (\ref{Tttan}) takes the form
\begin{equation}\label{Tttanap}
T=T_0e^{-\frac{DR_0^3}{2Ep}t^2} \ .
\end{equation}  
Now  looking on the time dependence of $T$
given in (\ref{Tttan}) or in its approximate
form (\ref{Tttanap})
we obtain following physical picture. 
For $D>0$ the tachyon is decreasing function
so that at same finite 
time it enters  the region where
the approximation of the large tachyon breaks down.
This time evolution of tachyon  describes the initial
stage of the process when an unstable D-brane
evolves from  the state 
close to its vacuum state to the state that can
be interpreted as emergence of the unstable Dp-brane.
On the other hand the case
when  $D<0$ corresponds to Dp-brane decay in
the process of the tachyon condensation.  

As the final example we will consider the
case when we split  $D$ and $E$   as $D=D_1+D_2, 
E=E_1+E_2$ 
where  $D_{1,2}$ and $E_{1,2}$ are related
to $T,R$ as
\begin{equation}\label{d12}
D_1+E_1t=-pTP_T \ ,
D_2+E_2t=-P_RR \ .
\end{equation}
Using the second relation in
(\ref{d12}) we  express $P_R$ as
a function of $R,E_2$ and $D_2$
and insert it into (\ref{dRT}). 
Then we obtain the differential equation
for $R$
\begin{equation}
\dot{R}=-\frac{(D_2+tE_2)R^3}
{\lambda E}
\end{equation}
that has the solution
\begin{equation}\label{Rtg}
\frac{\lambda}{R^2}=
\frac{2D_2}{\lambda E}t+
\frac{E_2}{E}t^2+\frac{\lambda}{R_0^2} \ .
\end{equation}
For $D_2>0$ the physical interpretation
of this solution is the same as in examples
studied above. On the
other hand for $D_2<0$ we obtain following
picture: Initially the radial mode grows
until it reaches its  turning point at 
$t_*=-\frac{D_2}{E_2}$ where $\dot{R}=0$.
Then Dp-brane starts to move towards
to the worldvolume of $N$ D3-branes. 

Let us now briefly discuss the time evolution
of tachyon. Inserting the first equation in
(\ref{d12}) into (\ref{dRT}) we get following
differential equation for $T$
\begin{equation}\label{Ttg}
\dot{T}=-\frac{(D_1+E_1t)TR^2}{
2\sqrt{\lambda}pE} \ .
\end{equation}
Inserting  (\ref{Rtg})
into the equation given above we
obtain the differential equation
for $T$ that has roughly following
form
\begin{equation}
\frac{dT}{T}=f(t)dt \ ,
\end{equation}
where $f(t)$ is a rational function that arises
by inserting (\ref{Rtg}) into 
(\ref{Ttg}). However in
order to obtain the rough
picture of the time   evolution of tachyon it is sufficient
to study the equation (\ref{Ttg})
without its explicit  integration. For 
$D_2>0$ the tachyon is always decreasing
and the tachyon evolution describes
the process of the emergence of 
Dp-brane from the state close to the
tachyon vacuum. On the other
hand for $D_2<0$ we
obtain that the tachyon firstly grows
until it reaches 
its turning point at $t_*=-\frac{D_1}{E_1}$
when $\dot{T}=0$ 
and then again decreases.  At any
case for nonzero $E_1$ there is not
any possibility to find solution describing
the decay of an unstable Dp-brane. 

Now we will discuss the dynamics of
unstable D0-brane. For $p=0$ the charge $D$
is equal to
\begin{equation}
D=-tE-P_RR 
\end{equation}
that allows us to express
$P_R$ as
\begin{equation}
P_R=-\frac{tE+D}{R}
\ .
\end{equation}
Then  the first equation in
(\ref{dRT}) gives
\begin{equation}\label{Rt}
\dot{R}=-R^3\frac{tE+D}{\lambda E}
\Rightarrow
\frac{\lambda}{R^2}-
\frac{\lambda}{R^2_0}=
t^2+\frac{2D}{E}t \ 
\end{equation}
for $t_0=0$. 
We see that an unstable D0-brane is
moving towards to the  D3-brane
 worldvolume  for $D\leq 0$. 
On the other hand for  $D<0$ 
the mode describing 
the radial position of D0-brane initially
grows then it reaches its turning point
at $t_*=-\frac{D}{E}$ 
and then D0-brane moves towards to
the stack of D3-branes. In order
to obtain relatively simple expressions
we will consider the case when
$D$ and angular momentum $P_{\theta}$
vanish.
 
Using the equation (\ref{Rt}) 
we can easily determine  the time dependence
of $T$. Firstly we express 
$P_T$ as function of $E,T$ and $R$
\begin{eqnarray}
P_T^2
=\frac{2\sqrt{\lambda}E^2}
{T^2R^2}-\frac{4\tau_p^2}
{T^4}-\frac{2t^2E^2}
{\sqrt{\lambda}T^2} \ . 
\nonumber \\
\end{eqnarray}
Then we insert this expression into
(\ref{dRT}) so that we get
\begin{eqnarray}\label{dotTo}
\dot{T}=\frac{P_TT^2R^2}{2\sqrt{\lambda}
E}=\pm\sqrt{
\frac{T^2R^2}{2\sqrt{\lambda}}
-\frac{\tau_p^2R^4}{\lambda E^2}-\frac{t^2T^2R^4}
{2\lambda^{3/2}}} \ . \nonumber \\
\end{eqnarray}
In fact it is very difficult to solve this equation
in the full generality. Then in order to
exhibit the main futures of the 
time evolution of tachyon it
is useful to perform 
some reasonable approximation in  
(\ref{dotTo}). First of
all, let us presume that 
$\frac{t^2R_0^2}{\lambda}\ll 1 \Rightarrow
t^2\ll \frac{\lambda}{R_0^2}$. Since by presumption
$\frac{\lambda}{R_0^2}\gg 1$ we see that
this condition is obeyed  almost for all $t$.
In this approximation we have
$\frac{\lambda}{R^2}\approx \frac{\lambda}
{R_0^2}$ so that the
equation 
(\ref{dotTo}) simplifies as
\begin{equation}
\dot{T}=\pm\sqrt{\frac{T^2R_0^2}
{2\sqrt{\lambda}}-\frac{\tau_p^2R_0^4}{\lambda
E^2}} 
\end{equation}
that has the solution
\begin{equation}
T==
T_0\pm\frac{\sqrt{2}\tau_p R_0}
{E\lambda^{1/4}}\cosh\left(\frac{\lambda^{1/4}}
{\sqrt{2}}
\frac{R_0}{\sqrt{\lambda}}t\right) \ .
\end{equation}
 As in the case
of the unstable D2-brane we obtain
two solutions corresponding to 
D0-brane decay  and the initial
stage of the emergence of  D0-brane
respectively. 
 However since the factor
in the function $\cosh(x)$ is proportional
to $\frac{R_0t}{\sqrt{\lambda}}\ll 1$ we
see that the tachyon is almost constant around
the interval under which the approximation
used above is valid which implies that the time
evolution is very slow. 

Let us now determine the time
evolution of tachyon for 
$\frac{t^2R_0^2}{\lambda} \approx  1$.
In the first approximation  we can
write $\frac{\lambda}{R^2}
\approx \frac{2\lambda}{R_0^2}$
and hence the differential equation for
$T$ takes the form
\begin{equation}
\dot{T}\approx \pm\sqrt{\frac{T^2R_0^2}
{8\sqrt{\lambda}}-\frac{\tau_p^2R_0^4}{4\lambda E^2}}
\end{equation}
with the solution
\begin{equation}
T\approx \pm\cosh\left(\frac{R_0}
{2\sqrt{2}\sqrt{\lambda}}t\right)
\end{equation}
that again implies that the tachyon is growing or
decreasing  very
slowly. 

Let us outline the result that
were obtained in this section. 
We have shown that
in order to find additional 
conserved
charge we should restrict to the case of $k=3$.
Then we have studied the dynamics of 
unstable D2 and D0-branes in
the background of $N$ D3-branes. 
We have seen that the dynamics of
the tachyon and radial mode is very reach.
In fact, we have argued that the tachyon
evolution describes either initial stage
of the emergence of unstable Dp-brane or
complete decay of this D-brane respectively
while it moves towards to the worldvolume
of $N$ D3-branes.  We have also seen
that the form of the time evolution
depends on the worldvolume of
unstable Dp-brane.

%%%%%%%%%%%%%%%%%%%%%%%%%%%%%%%%%%%%%%%%%%%%%
\section{Spatial dependent tachyon}\label{fourth}
In this section we will consider
the second example 
 of the tachyon condensation 
where we will presume that $T$ depends
on one  spatial coordinate on the non-BPS
Dp-brane worldvolume,
 say $x=\xi^1$, 
while the modes that describe embedding
of the unstable Dp-brane
in transverse space  are functions
of time. From As in the
case if the spatial
dependent tachyon condensation
in flat spacetime one can expect
that a codimension one D(p-1)-brane will
emerge that then   moves in the background
of $N$ Dk-branes. Let us check that
this guess is correct.

As is well known from the
study of the tachyon condensation in flat
spacetime \cite{Sen:2003tm}
 it  is useful to know the form of the stress
energy tensor $T_{\mu\nu}$
for the
modes living on the worldvolume
of non-BPS Dp-brane. In order to
find this stress energy tensor it
is useful to replace the flat
metric $\eta_{\mu\nu}$ 
that appears in the non-BPS Dp-brane
action with the general metric 
$g_{\mu\nu}$. Then the 
non-BPS Dp-brane action now takes
the form
\begin{eqnarray}
S=-\tau_p\int d^{p+1}\xi\sqrt{-g}
\sqrt{F}H_k^{\frac{k-p-4}{4}}\sqrt{\det\bA} \ ,
\nonumber \\
\bA^{\mu}_{\nu}=
  \delta^{\mu}_{\nu}+H_kg^{\mu\kappa}
\partial_{\kappa}X^m\partial_{\nu}X^m
+FH_k^{\frac{1}{2}}g^{\mu\kappa}
\partial_{\kappa}T\partial_{\nu}T
 \ . \nonumber \\
\end{eqnarray}
Consequently the stress energy tensor defined as
\begin{equation}
T_{\mu\nu}=-\frac{2}{\sqrt{-g}}\frac{\delta S}{\delta
g^{\mu\nu}}=-g_{\mu\nu}\mathcal{L}+2\frac{\delta 
\mathcal{L}}{\delta g^{\mu\nu}}
\end{equation}
is equal to
\begin{equation}
T_{\mu\nu}
=-g_{\mu\nu}\tau_p
\sqrt{F}H_k^{\frac{k-p-4}{4}}\sqrt{\det\bA}
+\tau_p
\sqrt{F}H_k^{\frac{k-p-4}{4}}
\tr\frac{\delta \bA}{\delta g^{\mu\nu}}
\bA^{-1}\sqrt{\det \bA}  \ , 
\end{equation}
where 
\begin{equation}
\tr \frac{\delta \bA}
{\delta g^{xy}}
\bA^{-1}=\frac{\delta \bA^{\mu}_{\kappa}}{\delta g^{xy}}
(\bA^{-1})^{\kappa}_{\mu}=
\left(
H_k\partial_yX^m
\partial_{\kappa}X^m+F
H_k^{\frac{1}{2}}
\partial_yT
\partial_{\kappa}T\right)  
(\bA^{-1})^{\kappa}_x  \ . 
\end{equation}
By definition the stress energy tensor
is conserved and hence its components
obey following differential equations
\begin{equation}
\partial_{\mu}\eta^{\mu\kappa}
T_{\kappa\nu}=0  \ . 
\end{equation}
For the spatial dependent 
tachyon and the time dependent
$X^m$'s the matrix $\bA$ is equal to
\begin{eqnarray}
\bA=\left(\begin{array}{ccc}
1-H_k\dot{X}^m\dot{X}^m & 0 & 0 \\
0 & 1+FH_k^{\frac{1}{2}}T'^2 & 0 \\
0 & 0 & \bI_{(p-1)\times (p-1)} \\
\end{array}\right) \  \nonumber \\
\end{eqnarray}
so that $\det \bA=(1-H_k\dot{X}^m\dot{X}^m)
(1+H_k^{\frac{1}{2}}FT'^2).$
Then the components of the stress energy tensor
are equal to
\begin{eqnarray}
T_{00}= \frac{\tau_p
\sqrt{F}H_k^{\frac{k-p-4}{4}}
\sqrt{1+FH_k^{\frac{1}{2}}T'^2}}
{\sqrt{1-H_k\dot{X}^m\dot{X}^m}} \ , 
T_{0i}=0 \ , i=1,\dots, p
\nonumber \\
T_{xx}=-\frac{\tau_p
\sqrt{F}H_k^{\frac{k-p-4}{4}}
\sqrt{1-H_k\dot{X}^m\dot{X}^m}}
{ \sqrt{1+FH_k^{\frac{1}{2}}T'^2}} 
\ , T_{xi}=0 \ , i=2,\dots, p
\nonumber \\
T_{ij}=-\delta_{ij}\tau_p
\sqrt{F}H_k^{\frac{k-p-4}{4}}
\sqrt{1-H_k\dot{X}^m\dot{X}^m}
\sqrt{1+FH_k^{\frac{1}{2}}T'^2} \ . 
\nonumber \\
\end{eqnarray}
In order to find
the lower dimensional D-brane
 we will closely follow 
\cite{Sen:2003tm}. Since for the
tachyon kink $T\rightarrow -\infty$ for 
$x\rightarrow -\infty$ and $T\rightarrow
\infty$ for $x\rightarrow \infty$  and
$\sqrt{F(T)}\rightarrow 0$ in this limit 
we have that $T_{xx}\rightarrow 0$ at
asymptotic $x$. Since now $T_{xx}$ is 
independent on $x$ we see that it must 
vanish for all $x$. This however implies that
we should have 
\begin{equation}
T=\pm \infty \ \mathrm{or}
\ \partial_x T=\pm \infty \ 
\mathrm{(or \  both)} \ 
\mathrm{for} \ \mathrm{all} \  x \ . 
\end{equation}
Clearly this solution is singular. Following
\cite{Sen:2003tm} we will show that this solution
has finite energy density that is localised
on codimension one subspace that
is expected to be D(p-1)-brane. To
see this let us consider the field configuration
\begin{equation}\label{anst}
T(x)=f(ax)
\end{equation}
where $f(u)$ satisfies 
\begin{equation}\label{ans}
f(-u)=-f(u) \ , 
g'(u)>0 \forall \ u \ , g'(u)=\sqrt{F(f(u))}
f'(u) \ , 
f(\pm \infty)=\pm \infty \ ,
\end{equation}
but is otherwise arbitrary function
of $u$. a is constant that we  should
take to $\infty$ at the end.  In this
case we have $T<0 $ for $x<0$ and
$T>0$ for $x>0$ and hence this
kink is singular as we expected.
Note also that the function $g(u)$
is related to $f(u)$ as
\begin{equation}
dg=\frac{1}{\sqrt{1+\frac{f^2}{2}}}df
\Rightarrow
\sinh\frac{g}{\sqrt{2}}=
\frac{f}{\sqrt{2}} \ . 
\end{equation}
Then for (\ref{anst}) $T_{xx}$ is equal
to
\begin{equation}
T_{xx}=-\frac{\tau_p
\sqrt{F(f(ax))}
H_k^{\frac{k-p-4}{4}}\sqrt{1-H_k\dot{X}^m\dot{X}^m}}
{ \sqrt{1+a^2F(f(ax))H_k^{\frac{1}{2}}f'^2(ax)}} \ .
\end{equation}
Clearly this component vanishes for all $x$ since
the numerator vanishes (except for $x=0)$ and
the denominator in the limit $a\rightarrow \infty$
is proportional to
\begin{equation}
\sqrt{1+H_kFT'^2}=
\sqrt{1+H_k a^2g'^2(ax)}
\sim a\rightarrow \infty  \ . 
\end{equation}
One can also easily check,
following \cite{Sen:2003tm}
that the ansatz (\ref{ans}) 
solves the equation
of motion for $T$ in the limit
$a\rightarrow \infty$.

Now using the ansatz (\ref{ans})
it is easy to find
other components of the stress energy tensor
\begin{eqnarray}
T_{00}(x)=\frac{\tau_pH_k^{\frac{k-p-4}{4}}
}{\sqrt{
1-H_k\dot{X}^m\dot{X}^m}}\sqrt{F}
\sqrt{1+H_k^{\frac{1}{2}}FT'^2}=
\nonumber \\
=\frac{\tau_pH_k^{\frac{k-p-4}{4}}
}{\sqrt{
1-H_k\dot{X}^m\dot{X}^m}}H_k^{\frac{1}{4}}
F(f(ax))
af'(ax) \ , \nonumber \\
T_{ij}(x)=-\delta_{ij}\tau_pH_k^{\frac{k-p-4}{2}}
\sqrt{1-H_k\dot{X}^m\dot{X}^m}
\sqrt{F}\sqrt{1+H_k^{\frac{1}{2}}F
T'^2}=\nonumber \\
=-\delta_{ij}\tau_pH_k^{\frac{k-p-4}{4}}
\sqrt{1-H_k\dot{X}^m\dot{X}^m}
H_k^{\frac{1}{4}}F(f(ax))
af'(ax)\nonumber \\
\end{eqnarray}
 in the limit $a\rightarrow \infty$.
Then the integrated $T_{00} \ ,
T_{ij}$ associated
with the codimension one solution are equal
to
\begin{eqnarray}
T^{kink}_{00}=
\int dx T_{00}=
\frac{\tau_pH_k^{\frac{k-(p-1)-4}{4}}}{\sqrt{
1-H_k\dot{X}^m\dot{X}^m}}
\int dx F(f(ax))af'(ax)=\nonumber \\
=\frac{\tau_pH_k^{\frac{k-(p-1)-4}{4}}
}{\sqrt{
1-H_k\dot{X}^m\dot{X}^m}}
\int dy F(y)dy
\nonumber \\
T^{kink}_{ij}=-\delta_{ij}
\tau_pH_k^{\frac{k-(p-1)-4}{2}}
\sqrt{1-H_k\dot{X}^m\dot{X}^m}
\int
dxF(f(ax))
af'(ax)=\nonumber \\
=
-\delta_{ij}
\tau_pH_k^{\frac{k-(p-1)-4}{2}}
\sqrt{1-H_k\dot{X}^m\dot{X}^m}
\int dy F(y)dy \ ,
\nonumber \\
\end{eqnarray}
where $y=f(ax)$. 
Thus 
$T^{kink}_{\alpha\beta} \ , \alpha , \beta=
0,2,\dots,p$ 
depend
on $F$ and not on the form of $f(u)$.
It is clear from the form of the function
$F$ that
most of the contribution is contained
in the finite range of $y$. In fact,
in the limit $a\rightarrow \infty$
the stress energy tensor $T^{kink}_{\alpha\beta}$
is localised on codimension one
D(p-1)-brane with the tension given
as 
\begin{equation}
T_{p-1}=\tau_p\int dy F=\tau_p
\int \frac{dy}{1+\frac{y^2}{2}}=
\sqrt{2}\pi \tau_p=\frac{(2\pi)}{(2\pi)^{
p+1}}=\frac{1}{(2\pi)^p}
\end{equation}
using the fact that 
$\tau_p$ is equal to
\begin{equation}
\tau_p=\frac{\sqrt{2}}
{(2\pi)^{p+1}}
\ .
 \end{equation}
In other words, the spatial dependent
condensation leads to the emergence of
D(p-1)-brane that is moving in the
Dk-brane background.
 Since this situation was 
analysed previously
in \cite{Burgess:2003mm,Panigrahi:2004qr}
we will not repeat it there.
In fact, the aim  this section 
was to show that one
can construct codimension one objects
on the worldvolume of  a non-BPS Dp-brane
even in  the nontrivial background produced
by stack of $N$ Dk-branes
\footnote{Before we conclude this section
we should say  
few words about the fate of the WZ term in the
non-BPS Dp-brane action in
case of spatial dependent tachyon.
In this case  this term 
is approximately equal to
$\sim \int T'dx\wedge C_{p-1}(X^m)$. 
Its contribution  to the
tachyon equation of motion is 
proportional to $\int dx \frac{d\delta T}{dx}C
\Rightarrow -\int dx \delta T\frac{d}{dx}C(X^m) 
=0$ since by presumption 
$X^m$ are not functions of $x$.}.
%%%%%%%%%%%%%%%%%%%%%%%%%%%%%%%%%%%%%%
%%%%%%%conclussion %%%%%%%%%%%%%%%%%%%%%%%
%%%%%%%%%%%%%%%%%%%%%%%%%%%%%%%%%%%%%%
\section{Conclusion}\label{fifth}
This paper was devoted to the study
of the dynamics of a
non-BPS Dp-brane in
the background of $N$ Dk-branes, 
where 
$k>p$. It can be 
also considered as an
extension of the previous paper 
\cite{Kluson:2004xc} where the dynamics
of a non-BPS Dp-brane in the background
of $N$ NS5-branes was studied. 
We were mainly interested in the time
evolution  of the tachyon and
radion mode on the worldvolume of the
non-BPS Dp-brane. We have shown 
that 
the main properties of the dynamics of
the  unstable Dp-brane 
strongly depends on dimensions of
the background Dk-branes and the probe Dp-brane. 
Generally the time dependent
 tachyon evolution either 
reduces or increases 
  the initial tension of  a non-BPS Dp-brane 
and hence one can expect that
the tachyon condensation would
have profound impact on the motion
of an unstable  Dp-brane 
in the curved spacetime. 
To support this claim we have restricted
ourselves to the study of the
unstable  D-brane dynamics  
 in the region of  
the field theory space 
where $T$ is large  
and $\frac{\lambda}{R^{7-k}}\gg 1$.  
We have then argued that 
for the background of $N$ D3-branes 
a new symmetry on the 
worldvolume of the unstable
Dp-brane emerges. Using now
the conserved  charge that
is generator of this symmetry 
 we were able to 
sketch the rough picture of the
 evolution of $T$ and $R$
that live 
on the worldvolume of non-BPS
D2 and D0-branes. 

We mean that the fact that the
form of the tachyon condensation
strongly depends on dimensions of
background Dk-branes and  a probe
Dp-brane is related to the fact
the charge $D$ is defined for the
background of $N$ D3-branes only. 
It would be certainly very interesting
to find 
general form of the transformation
of the worldvolume fields  
that will leave the effective action
invariant. We hope that the 
possible existence of this symmetry
could be helpful for better understanding
of the  geometrical
origin of the tachyon conjecture. 
 We  mean   that 
this symmetry could be
related to the  generalised symmetry 
discussed in
\cite{Jevicki:1998ub,Jevicki:1998yr}.
We plan to return to this problem
in the future work.
\\
\\
{\bf Acknowledgement}

This work was supported by the
Czech Ministry of Education under Contract No.
MSM 0021622409.

%%%%%%%%%%%%%%%%%%%%%%%%%%%%%%%%%%%
%%%%%%% Thebibligraphy %%%%%%%%%%%%%%%%%%%%%
%%%%%%%%%%%%%%%%%%%%%%%%%%%%%%%%%%%%%

\end{document}